\documentclass[aps,prb,reprint,superscriptaddress]{revtex4-2}

\usepackage{amsmath,amssymb}
\usepackage{graphicx}

\AtBeginDocument{%
  \setlength{\abovedisplayskip}{8pt plus 2pt minus 2pt}%
  \setlength{\belowdisplayskip}{8pt plus 2pt minus 2pt}%
  \setlength{\abovedisplayshortskip}{5pt plus 2pt minus 2pt}%
  \setlength{\belowdisplayshortskip}{6pt plus 2pt minus 2pt}%
}

\begin{document}

\title{Acoustic Phonon Dynamics in Co-Doped BaFe$_2$As$_2$ Thin Film}

\author{Alexander Bartenev}
\affiliation{Department of Physics, University of Puerto Rico, Mayag\"uez, Puerto Rico 00681, USA}
\author{Roman Kolodka}
\affiliation{Department of Physics, University of Puerto Rico, Mayag\"uez, Puerto Rico 00681, USA}
\author{Larry Theran}
\affiliation{Department of Physics, University of Puerto Rico, Mayag\"uez, Puerto Rico 00681, USA}
\author{Ki-Tae Eom}
\affiliation{Department of Materials Science and Engineering, University of Wisconsin--Madison, Madison, WI 53706, USA}
\author{Jong-Hoon Kang}
\affiliation{Department of Materials Science and Engineering, University of Wisconsin--Madison, Madison, WI 53706, USA}
\author{Jason Kawasaki}
\affiliation{Department of Materials Science and Engineering, University of Wisconsin--Madison, Madison, WI 53706, USA}
\author{Chang-Beom Eom}
\affiliation{Department of Materials Science and Engineering, University of Wisconsin--Madison, Madison, WI 53706, USA}
\author{Sergiy Lysenko}
\email[Corresponding author: ]{sergiy.lysenko@upr.edu}
\affiliation{Department of Physics, University of Puerto Rico, Mayag\"uez, Puerto Rico 00681, USA}

\begin{abstract}
Pump-probe reflectivity reveals coherent acoustic oscillations at 33 and 8.2~GHz in a Ba(Fe$_{0.92}$Co$_{0.08}$)$_2$As$_2$ thin film. The acoustic response was analyzed using a modified logistic-function model, suggesting a temporal redistribution of coherent acoustic energy consistent with anharmonic decay of the higher-frequency mode into the lower-frequency mode.

\end{abstract}

\maketitle

\section{Introduction}

In iron-based superconductors, electronic nematicity lowers the rotational symmetry of the electronic system and couples strongly to lattice strain~\cite{Bohmer2022}. Recent work has shown that anisotropic strain can strongly suppress the superconducting transition in Co-doped BaFe$_2$As$_2$, emphasizing that the lattice is not a passive background but an active control parameter for the electronic phase diagram~\cite{Malinowski2020}. Picosecond ultrasonics probes the lattice response by launching coherent strain pulses~\cite{Suzuki2021}. The resulting transient reflectivity oscillations encode sound velocity, attenuation, and interface effects~\cite{Thomsen1984}. In BaFe$_2$As$_2$ thin films, coherent longitudinal acoustic phonons have recently been used to determine the out-of-plane elastic stiffness and to show that acoustic mode frequencies are sensitive to film thickness, intrinsic strain, and orthorhombicity~\cite{Cheng2023}. In this work, we use pump-probe reflectivity to study coherent acoustic phonons and their decay in a Ba(Fe$_{0.92}$Co$_{0.08}$)$_2$As$_2$ thin film. 

\section{Experiment}

Pump-probe measurements were performed in reflection geometry using a Spectra-Physics Ti:sapphire femtosecond laser system. Laser pulses with 35~fs duration and a central wavelength of  $\lambda$=800 nm were produced by a regenerative amplifier at a 1 kHz repetition rate. Pump and probe beams were spatially overlapped on the sample. The pump fluence was $F=5$~mJ/cm$^2$. The probe was attenuated with a neutral-density filter to minimize probe-induced excitation. The temperature was regulated in a closed-cycle cryostat down to 7.5~K with 0.1~K precision. An 80-nm-thick Ba(Fe$_{0.92}$Co$_{0.08}$)$_2$As$_2$ film was grown by pulsed-laser deposition on a single-crystal (La,Sr)(Al,Ta)O$_3$ (LSAT) substrate with a 40-nm-thick epitaxial SrTiO$_3$ (STO) buffer layer.

\section{Results and Discussion}

The coherent acoustic phonon contribution to the transient reflectivity signal can be described as a sum of oscillatory modes with time-dependent amplitudes~\cite{Bartenev2023}:
\begin{equation}
\frac{\Delta R}{R}(t)
=\sum_i R_i(t)=
\sum_i L_i(t)\sin\left(\omega_i t+\phi_i\right),
\label{eq:cap-signal}
\end{equation}
where $\omega_i$ is the angular frequency of the $i$-th phonon mode, $\phi_i$ is its initial phase, and $L_i(t)$ is a time-dependent modified logistic function that accounts for the generation, duration, and decay of each phonon mode:
\begin{equation}
L_i(t)
=
\frac{A_i}
{
1+\exp\left[-k_{g,i}(t-t_{g,i})\right]
+\exp\left[k_{d,i}(t-t_{d,i})\right]
}.
\label{eq:modified-logistic}
\end{equation}
Here, $A_i$ is the amplitude, $k_{g,i}$ and $k_{d,i}$ are the generation and decay rates, respectively, and $t_{g,i}$ and $t_{d,i}$ are temporal offset parameters associated with the growth and decay of the mode. The full width at half maximum (FWHM) of $L_i(t)$ can be used as an estimate of the phonon-mode lifetime. The fit parameters are presented in Table~\ref{tab:modified-logistic-parameters}.

\begin{table*}
\centering
\caption{Parameters of the modified logistic-function model.}
\label{tab:modified-logistic-parameters}
\begin{tabular}{c c c c c c c c}
\hline
Mode & $A_i$ & $k_{g,i}$ (ps$^{-1}$) & $k_{d,i}$ (ps$^{-1}$) & $t_{g,i}$ (ps) & $t_{d,i}$ (ps) & $\omega_i$ (rad/ps) & $\phi_i$ (rad) \\
\hline
$R_1(t)$ & $1.573\times 10^{-3}$ & 0.111 & 0.048 & 35.5 & 94.0 & 0.2074 & -0.08 \\
$R_2(t)$ & $9.55\times 10^{-4}$  & 0.111 & 0.052 & 75.5 & 237.6 & 0.0518 & 2.75 \\
\hline
\end{tabular}
\end{table*}

\begin{figure*}[t]
\centering
\includegraphics[width=0.95\textwidth]{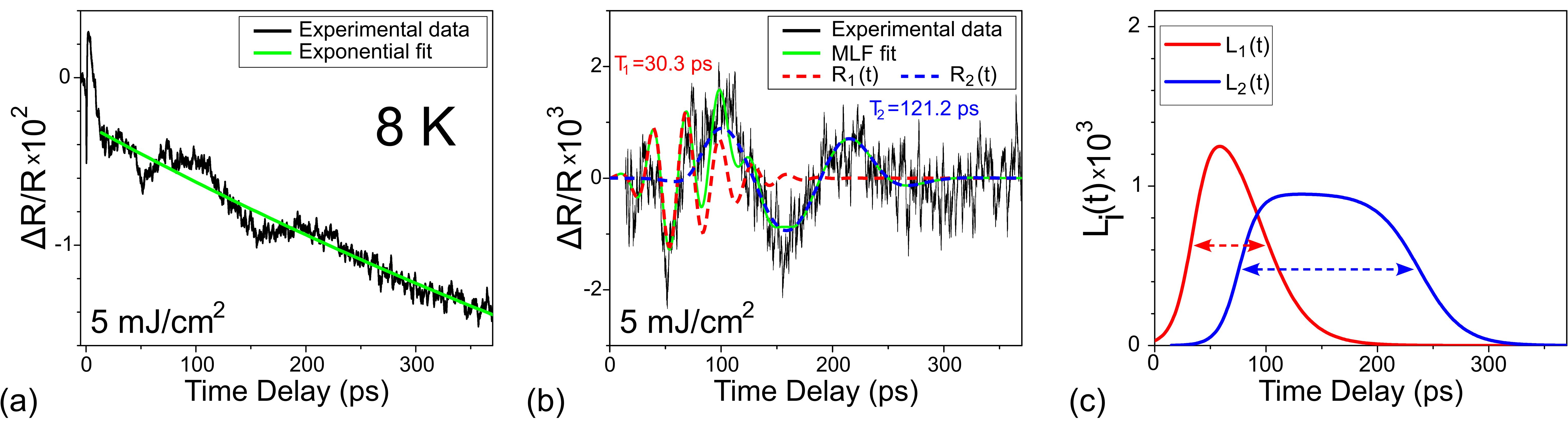}
\caption{\label{fig:fig1} (a) Transient reflectivity signal of Ba(Fe$_{0.92}$Co$_{0.08}$)$_2$As$_2$ thin film within 380~ps after photoexcitation. The solid line shows the exponential thermal background. (b) Residual acoustic response fitted with the modified logistic-function (MLF) model. (c) Modified logistic functions corresponding to the two acoustic phonon modes. The dashed arrows indicate the full widths at half maximum (FWHM). }
\end{figure*}

Photoexcitation generates a transient stress profile that launches a coherent acoustic wave [Fig.~\ref{fig:fig1}(a,b)]~\cite{Thomsen1984,Suzuki2021,Bartenev2023}. The resulting oscillatory component becomes apparent after approximately 20~ps. The acoustic response was fit using the modified logistic-function model shown in Eqs.~\eqref{eq:cap-signal} and~\eqref{eq:modified-logistic}. This response indicates a multicomponent acoustic signal, suggesting acoustic phonon decay driven by anharmonic interactions. The response contained two coherent acoustic components with frequencies of $f_1 = 33$~GHz and $f_2 = 8.2$~GHz.

The fitted logistic components have characteristic widths $FWHM_{L_1}=70.6$~ps and $FWHM_{L_2}=162.4$~ps, estimated as the full-width-at-half-maximum (FWHM) from the corresponding logistic functions [Fig.~\ref{fig:fig1}(c)]. These widths show that the high-frequency $f_1$ mode is noticeably shorter-lived than the low-frequency $f_2$ mode. The delayed onset of the lower-frequency mode together with the shorter lifetime of the higher-frequency mode is consistent with a temporal redistribution of coherent acoustic energy from the 33 GHz mode to the 8.2 GHz mode. Thus, the obtained temporal profiles $L_1(t)$ and $L_2(t)$ suggest anharmonic decay dynamics of $f_1$ into the lower-frequency $f_2$ mode.

As the acoustic pulse penetrates deeper into the film and reflects at the film-substrate interface, it loses coherence and decays, leading to low-frequency oscillations. In this case, the relationship between the speed of sound and the period $T_S$ is $v_s = 4d/T_S$~\cite{Thomsen1984}, where $d$=80 nm is the film thickness and $T_S$=$T_2$ is the oscillation period of the $R_2$ component. The obtained speed of sound, $v_s = 2640$~m/s, is found to be lower than the reported sound speed in parent BaFe$_2$As$_2$ thin films (3100--3400~m/s~\cite{Suzuki2021, Cheng2023}). Since the sound speed and longitudinal acoustic mode frequency are sensitive to lattice parameters, film thickness, and orthorhombicity in BaFe$_2$As$_2$~\cite{Cheng2023}, the reduced value of $v_s$ may be attributed to elastic softening caused by Co substitution and thin-film strain.

The laser pulse excites a longitudinal strain wave with the initial frequency of $f_1$, associated with the acoustic pulse passing through the film thickness, corresponding to the acoustic length of the film. This coherent longitudinal acoustic thickness mode rapidly loses coherence due to energy transfer to the substrate, dephasing at the substrate/film interface, and possibly anharmonic coupling with other acoustic modes. The propagation time of the acoustic pulse through the 80 nm film is $\sim$30 ps. Taking into account that the acoustic response becomes observable only after a $\sim$20 ps delay, the $\sim$30 ps propagation time is in good agreement with the onset of excitation of the $f_2$ echo mode shown in Fig.~\ref{fig:fig1}(c) and likely results from the interaction of the $f_1$ mode with the substrate. The approximate relation $f_1 \simeq 4f_2$ is compatible with a redistribution of coherent acoustic energy between longitudinal thickness modes, possibly mediated by nonlinear internal resonance.

\vspace{1pt}
\smallskip 

\section*{Acknowledgments}

The authors gratefully acknowledge support from the U.S. Army Research Office, accomplished under Grant Number W911NF-25-1-0122, and from the National Science Foundation, Awards \#2425113 and \#1905691. C.-B.E. acknowledges support for this research through a Vannevar Bush Faculty Fellowship (ONR N00014-20-1-2844), and the Gordon and Betty Moore Foundation's EPiQS Initiative, Grant GBMF9065. Thin film synthesis and transport measurements at the University of Wisconsin--Madison were supported by the U.S. Department of Energy (DOE), Office of Science, Office of Basic Energy Sciences (BES), under award number DE-FG02-06ER46327. The authors gratefully acknowledge partial support of this research by NSF through the University of Wisconsin Materials Research Science and Engineering Center (DMR-2309000).


\begin{thebibliography}{99}

\bibitem{Bohmer2022} A. E. B\"ohmer, J.-H. Chu, S. Lederer, and M. Yi, ``Nematicity and nematic fluctuations in iron-based superconductors,'' Nat. Phys.~\textbf{18}, 1412--1419 (2022).

\bibitem{Malinowski2020} P. Malinowski, Q. Jiang, J. J. Sanchez, J. Mutch, Z. Liu, P. Went, J. Liu, P. J. Ryan, J.-W. Kim, and J.-H. Chu, ``Suppression of superconductivity by anisotropic strain near a nematic quantum critical point,'' Nat. Phys.~\textbf{16}, 1189--1193 (2020).

\bibitem{Suzuki2021} T. Suzuki, Y. Kubota, A. Nakamura, T. Shimojima, K. Takubo, S. Ito, K. Yamamoto, S. Michimae, H. Sato, H. Hiramatsu, H. Hosono, T. Togashi, M. Yabashi, H. Wadati, I. Matsuda, S. Shin, and K. Okazaki, ``Ultrafast Optical Stress on BaFe$_2$As$_2$,'' Phys. Rev. Res.~\textbf{3}, 033222 (2021).

\bibitem{Thomsen1984} C. Thomsen, J. Strait, Z. Vardeny, H. J. Maris, J. Tauc, and J. J. Hauser, ``Coherent phonon generation and detection by picosecond light pulses,'' Phys. Rev. Lett.~\textbf{53}, 989--992 (1984).

\bibitem{Cheng2023} D. Cheng, B. Song, J.-H. Kang, C. Sundahl, A. L. Edgeton, L. Luo, J.-M. Park, Y. G. Collantes, E. E. Hellstrom, M. Mootz, I. E. Perakis, C.-B. Eom, and J. Wang, ``Study of elastic and structural properties of BaFe$_2$As$_2$ ultrathin film using picosecond ultrasonics,'' Materials~\textbf{16}, 7031 (2023).

\bibitem{Bartenev2023} A. Bartenev, R. Kolodka, C. Verbel, M. Lozano, F. Fernandez, A. Rua, and S. Lysenko, ``Nonequilibrium carrier dynamics in FeSe$_{0.8}$Te$_{0.2}$,'' MRS Adv.~\textbf{8}, 167--172 (2023).

\end{thebibliography}
\end{document}